\documentclass[11pt]{article}
\usepackage
{
        nicefrac
}

\usepackage{fullpage}
\usepackage{floatrow}
\usepackage{amsthm,amsmath,amssymb}
\usepackage{natbib,hyperref}

\newcommand{\rclose}{\rho_{\mathrm{core}}}
\newcommand{\rfar}{\rho_{\mathrm{inter}}}

\DeclareMathOperator{\core}{core}

\DeclareMathOperator {\cost}{cost}

\DeclareMathOperator {\optcost}{opt-cost}

\DeclareMathOperator {\Var}  {Var}
\newcommand{\lgap}{\lambda_{gap}}

\newcommand {\hEN}{{\hat E_{-}}}
\newcommand {\hEE}{{\hat E_{+}}}
\newcommand{\hSDP}{\widehat{SDP}}
\newcommand{\hf}{{\hat f}}
\newcommand{\hcalI}{{\hat\calI}}
\newcommand{\Efix}{E_{\mathrm{flip}}}

\newcommand{\rsclose}{\rho^2_{\mathrm{core}}}

\newcommand{\rsavg}{\rho^2_{\mathrm{avg}}}

\newcommand {\bbR}    {\mathbb{R}}

\newcommand {\calW}   {{\cal{W}}}

\newcommand {\calP}   {{\cal{P}}}
\newcommand {\calF}   {{\cal{F}}}
\newcommand {\calZ}   {{\cal{Z}}}

\newcommand {\calI}   {{\cal{I}}}
\newcommand {\calC}   {{\cal{C}}}

\newcommand {\EN}{E_{-}}
\newcommand {\EE}{E_{+}}

\newcommand {\Set}   [1] {\left\{ #1 \right\}}

\newcommand {\uv}       {\langle \bar u , \bar v \rangle}

\newcommand {\Exp}       {\mathbb{E}}

\newcommand {\iprod} [2] {\langle #1, #2 \rangle}

\newcommand{\given}{\mid}

\newcommand{\symdiff}{\triangle}

\newcommand{\textproblem}[1]{{#1}} 

\newcommand{\eps}{\varepsilon}

\newtheorem{theorem}{Theorem}[section]
\newtheorem{lemma}[theorem]{Lemma}

\newtheorem{corollary}[theorem]{Corollary}
\newtheorem{definition}[theorem]{Definition}

\newtheorem{remark}{Remark}[section]

\newtheorem{assumptions}[theorem]{Assumptions}

\usepackage{times}

\begin{document}
\title{Correlation Clustering with Noisy Partial Information}



\author{Konstantin Makarychev\\Microsoft Research
\and Yury Makarychev\\TTIC \thanks{Supported by NSF CAREER award CCF-1150062 and NSF award IIS-1302662.}
\and Aravindan Vijayaraghavan\\Courant Institute, NYU \thanks{Supported by the Simons Collaboration on Algorithms and Geometry.}}
\date{}

\maketitle

\begin{abstract}
In this paper, we propose and study a semi-random model for the Correlation Clustering problem on arbitrary graphs $G$. We give two approximation algorithms for Correlation Clustering instances
from this model. The first algorithm finds a solution of value $(1+ \delta)\optcost + O_{\delta}(n\log^3 n)$ with high probability, where $\optcost$ is the value of the optimal solution (for every $\delta > 0$). The second algorithm finds the ground truth clustering with an arbitrarily small classification error $\eta$ (under some additional assumptions on the instance).
\end{abstract}

\section{Introduction} \label{sec:intro}

One of the most commonly used algorithmic tools in data analysis and machine learning is clustering -- partitioning a corpus of data into groups based on similarity. The data observed in several application domains -- e.g., protein-protein interaction data, links between web pages, and social ties on social networks  -- carry relational information between pairs of nodes, which can be represented using a graph. Clustering based on relational information can reveal important structural information such as functional groups of proteins~\citep{bio2, bio3}, communities on web and social networks~\citep{Fortunato,newman}, and can be used for predictive tasks such as link prediction~\citep{taskar}.

Correlation clustering tackles this problem of clustering objects when we are given qualitative information about the similarity or dissimilarity between some pairs of these objects.
This qualitative information is represented in the form of a graph $G(V,E,c)$ in which edges $E$ are labeled with signs $\Set{+,-}$; we denote the set of `$+$' edges by $\EE$ and the set of `$-$' edges by $E_{-}$. Each edge $(u,v)$ in $\EE$ indicates that $u$ and $v$ are similar, and each edge
$(u,v)\in \EN$ indicates that $u$ and $v$ are dissimilar; the cost $c(u,v)$ of the edge shows the amount of similarity or dissimilarity between $u$ and $v$.\footnote{One can also think of the instance as a graph $G(V,E,c)$ with the edge costs $c: E \rightarrow [-1,1]$. If $c(u,v) > 0$ then $(u,v) \in \EE$, and If $c(u,v) < 0$ then $(u,v)\in \EN$.} In the ideal case, this qualitative information is consistent with the intended (``ground truth'') clustering. However, the qualitative information may  be noisy due to errors in the observations.  Hence, the goal is to find a partition $\calP$ of $G$
that minimizes the cost of inconsistent edges:
$$ \min_{\calP} \sum_{(u,v)\in \EE: \calP(u)\neq \calP(v)} c(u,v) +
\sum_{(u,v)\in \EN: \calP(u) = \calP(v)} c(u,v),$$
where $\calP(u)$ denotes the cluster that contains the vertex $u$.
The objective captures  the cost of inconsistent edges -- cut edges in $\EE$ and uncut edges in $\EN$.
(For a partition $\calP$, we say that an edge $(u,v)\in E$ is consistent with $\calP$ if either $(u,v)\in E_{+}$ and $\calP(u) = \calP(v)$ or $(u,v)\in E_{-}$ and $\calP(u) \neq \calP(v)$.)

 Note that the underlying graph $G(V,E)$ can be reasonably sparse; this is desirable since collecting pairwise information can be expensive. One important feature of correlation clustering is that  it, unlike most other clustering problems, allows us  not to specify the number of clusters.
Hence, it is particularly
useful when we have no prior knowledge of the number of clusters that the data divides into.

Correlation clustering also comes up naturally in MAP inference in graphical models and structured prediction tasks for such tasks as image segmentation, parts-of-speech tagging and dependency parsing in natural language processing~\citep{Nowozkin11, noahsmith}. In structured prediction, we are given some observations as input (e.g., image data, sentences), and the goal is to predict a labeling $\mathbf{x} \in \mathcal{X}$ that encodes the high-level information that we would like to infer. For instance, in image segmentation, the variables $x \in \Set{0,1}^n$ indicate whether each pixel is in the foreground or background. This is naturally modeled as a Correlation Clustering instance on the set of pixels  (with 2 clusters), where edges connect adjacent pixels, and the costs (with signs) are set based on the similarity or dissimilarity of the corresponding pixels in the given image. The clusters in these inference problems  then consist of the sets of variables that receive the same assignment in the MAP solution.
Correlation clustering is also used in the context of consensus clustering and agnostic learning.

Correlation clustering was introduced in~\citep{BBC}, and implicitly in~\citep{BSY} as `Cluster Editing'.
The problem is APX-hard even on complete graphs\footnote{This rules out $(1+\epsilon)$ factor approximations for some small constant $\epsilon>0$.} (when we are given the similarity information for every pair of objects) \citep{CGW}.  The state-of-the-art approximation algorithm~\citep{CGW,DEFI} achieves an $O(\log n)$ approximation for minimizing disagreements in the worst-case. Furthermore, there is a gap-preserving reduction from the classic Minimum Multicut problem \citep{CGW,DEFI}, for which the current state-of-the-art algorithm gives a $\Theta(\log n)$ factor approximation \citep{GVY}.
The complementary objective of maximizing agreements is easier from the approximability standpoint, and a 0.766 factor approximation is known~\citep{CGW,Swamy}. For the special case of complete graphs (with unit costs on edges), small constant factor approximations have been obtained  in a series of works~\citep{BBC,ACN,CMSY}. Instances of Correlation Clustering on complete graphs that satisfy the notion of approximation stability were considered in \citep{BB09}. To summarize, despite our best efforts, we only know logarithmic factor approximation algorithms for Correlation Clustering; moreover, we cannot get a constant factor approximation for worst-case instances if the Unique Games Conjecture is true.

However, our primary interest in solving Correlation Clustering comes from its numerous applications, and  the instances that we encounter in these applications are not worst-case instances. This motivates the study of the average-case complexity of the problem and raises the following question:
\begin{quote}
Can we design algorithms with better provable guarantees for realistic
average-case models of \textproblem{Correlation Clustering}?
\end{quote}

Several natural average-case models of Correlation Clustering have been studied previously. 
\cite{BSY} consider a model in which we start with a ground-truth clustering -- an arbitrary partitioning of the vertices -- of a complete graph. Initially, edges inside clusters of the ground truth solution are labeled `+' and edges between clusters are labeled `-'. We flip the label of each edge (change `$+$' to `$-$' and `$-$' to `$+$') with probability $\varepsilon$ independently at random and obtain a Correlation Clustering instance (the flipped edges model the noisy observations) . In fact, this average-case model was also studied in the work \citep{BBC} that introduced the problem of Correlation Clustering. 
\citeauthor{MS10} consider a generalization of this model where there is an adversary: for each edge, we keep the initial label with probability $(1-\varepsilon)$,
 and we let  the adversary decide whether to flip the edge label or not with probability $\varepsilon$.
The major drawback of these models is that they only consider the case of complete graphs, i.e. they require that the Correlation Clustering instance contains similarity
information for \emph{every} pair of nodes.
 \citeauthor{Chenetal} extended the model of \citep{BSY} from complete graphs to sparser Erdos--Renyi random graphs. In their model, the underlying unlabeled graph $G(V,E)$ comes from an Erd\"os--Renyi random graph (of edge probability $p$), and as in \citep{BSY}, the label of each edge is set (independently) to be consistent with the ground truth clustering with probability $1-\varepsilon$ and inconsistent with probability $\varepsilon$. 
 
While these average-case models are natural, they are unrealistic in practice since most real-world graphs are neither dense nor captured by  Erd\"os--Renyi distributions. 
For instance, real-world graphs in community detection have many structural properties (presence of large cliques, large clustering coefficients, heavy-tailed degree distribution) that are not exhibited by graphs that are generated by Erd\"os--Renyi models \citep{pref,KRRT}. Graphs that come up in computer vision applications are sparse with grid-like structure~\citep{YIF12}. Further, these models assume that every pair of vertices have the same amount of similarity or dissimilarity (all costs are unit). Our semi-random model tries to address these issues by assuming very little about the observations -- the underlying unlabeled graph $G(V,E)$  -- and allowing non-uniform costs.

\subsection{Our Semi-random Model}\label{sec:semirandom-informal}
In this paper, we propose and study a new semi-random model for generating general instances of Correlation Clustering, which we
believe captures many properties of real world instances. %
It generalizes the model of  
\cite{MS10} to arbitrary graphs $G(V,E,c)$ with costs.
A semi-random instance $\Set{G(V,E,c), (\EE, E_{-})}$ is generated as follows:
\begin{enumerate}
\item The adversary chooses an undirected graph $G(V,E,c)$ and a partition $\calP^*$ of the vertex set $V$ (referred to as the planted clustering or ground truth clustering).
\item Every edge is $E$ is included in set $E_R$ independently with probability $\varepsilon$.
\item Every edge $(u,v) \in E \setminus E_R$ with $u$ and $v$ in the same cluster of $\calP^*$  is included in $\EE$, and every edge $(u,v) \in E\setminus E_R$, with $u$ and $v$ in different clusters of $\calP^*$ is included in $\EN$.
\item The adversary adds every edge from $E_R$ either to $\EE$ or to $E_{-}$ (but not to both sets).
\end{enumerate}
This model can be further generalized to an adaptive semi-random model as described in Section~\ref{sec:adaptive}.

%

\subsection{Our Results}
We develop two algorithms for semi-random instances of Correlation Clustering. The first algorithm gives a polynomial-time approximation scheme (PTAS) for instances from our semi-random model. The second algorithm recovers the planted partition with a small classification error $\eta$.
\begin{theorem}\label{thm:main}
For every $\delta > 0$,  there is a polynomial-time algorithm that given a semi-random
instance $\Set{G(V,E,c),(\EE,E_{-})}$ of Correlation Clustering (with noise probability $\eps < 1/4$), finds
a clustering that has
disagreement cost $(1+\delta) \optcost + O((1-2\varepsilon)^{-4} \delta^{-3} n \log^3 n)$ w.h.p. over the randomness in the instance, where $\optcost$ is the cost of disagreements of the optimal solution for the instance.
\end{theorem}
The approximation additive term is much smaller than the cost of the planted solution
if the average degree $\Delta \gg \varepsilon^{-1}\mathrm{polylog}\ n$.
Note that we compare the performance of our algorithm with the cost of the \textit{optimal} solution. Further, these guarantees hold even in a more general adaptive semi-random model that is described in Section~\ref{sec:adaptive}.


The above result gives a good approximation guarantee with respect to the objective. \emph{But what about recovering the ground truth clustering?} Our semi-random model is too general to allow recovery. For instance, there could be large disconnected pieces inside some clusters of $G$, or there could be no edges between some clusters --- in both cases, recovery is statistically impossible. Hence, we need some additional conditions for approximate recovery in our model, that guarantee at the very least that the ground truth clustering is uniquely optimal (in a robust sense).

Our first assumption is that there is mild {\em expansion inside clusters} --- this connectivity assumption prevents large pieces inside clusters that are almost disconnected, which might get separated in an almost optimal clustering. The second and third assumptions are that there are enough edges from vertices in one cluster to other clusters, to prevent these clusters (or parts of them) from coalescing in near-optimal clusterings. Finally, we assume (approximate) regularity in degrees inside clusters, since it is hard to correctly classify vertices with very few edges incident on them. These assumptions are described formally in Assumptions~\ref{assumptions:recovery}. We now informally describe the algorithmic guarantees for approximate recovery:

\begin{theorem}
\label{thm:recovery_algorithm}
There exists a polynomial-time algorithm that given a semi-random instance \\
 $\calI = \Set{G = (V, E, c),(\EE, \EN)}$ satisfying mild expansion inside clusters, regularity and inter-cluster density conditions (see Assumptions~\ref{assumptions:recovery} for details)
 finds a partition $\cal P$ with classification error at most $4\eta$ w.h.p. over the randomness in the instance, where
\begin{equation}\label{eq:eta-def}
\eta= \frac{C_2}{1-2\varepsilon} \left( \frac{n\log n}{\cost(E)}\right)^{1/12} \cdot \left(\frac{1}{\beta\lgap}\right)^{1/2}.
\end{equation}
\end{theorem}

Our algorithm outputs a clustering such that only $O(\eta n)$ vertices are misclassified (up to a renaming of the clusters).  We note that the expansion and regularity assumptions  are satisfied by Erd\"os--Renyi graphs: for instance, such random graphs have strong expansion both inside and between clusters ($\lambda_{\text{gap}}=1-o(1)$) and have strong concentration of degrees. Our assumptions for recovery are soft: if there is bad expansion inside clusters ($\lambda_{\text{gap}}$ is small), or if there are not sufficient edges between vertices in different clusters, we just need more observations (edges) to approximately recover the clusters. We note that the regularity conditions in Assumptions~\ref{assumptions:recovery} are more for convenience and may be significantly relaxed. In particular, the same algorithm and analysis works even when the degrees are approximately regular (up to poly-logarithmic factors, for example) --- this irregularity just appears in equation~(\ref{eq:eta-def}) as an extra multiplicative factor. We defer these details to the journal version of our paper.


\subsection{Related Work on Semi-random Models}
Over the last two decades, there has been extensive research on average-case complexity of many important combinatorial
optimization problems. 
Semi-random instances typically allow much more structure then completely random instances.
Research on semi-random models was initiated by 
\citep{BS}, who introduced and investigated semi-ran\-dom models for $k$-coloring.
Semi-random models have also been studied for graph partitioning problems~\citep{FKil,Yudong,MMV,MMV2}, Independent Set~\citep{FKil}, Maximum Clique~\citep{FKra}, Unique Games~\citep{KMM}, and other problems.
Most related to our work, both in the nature of the model and in the techniques used, is a recent result of~\citep{MMVfas} 
 on semi-random instances of Minimum Feedback Arc Set. While the techniques used in both papers are conceptually similar, the semidefinite (SDP) relaxation for Correlation Clustering that we use in this paper is very different from the SDP relaxation for Minimum Feedback Arc Set used in~\citep{MMVfas}. Further, we get a \text{true}
$1+\delta$ approximation scheme (with an extra additive approximation term). This is in contrast to previous semi-random model results \citep{MMV, MMVfas}, which compare the cost of the solution that the algorithm finds to the cost of the planted solution. Moreover,  this work gives not only a PTAS for the problem, but also a simple algorithm for recovery the ground truth solution.

\citeauthor{MS10} recently considered a semi-random model for \textproblem{Correlation Clustering} on complete graphs with unit edge costs. Later, \citeauthor{ES} conducted an empirical evaluation of algorithms for the complete graph setting. \cite{Chenetal} extended the average-case model of Correlation Clustering to sparser Erd\"os--Renyi graphs. Very recently, \cite{GRSY14} considered a semi-random model for Correlation Clustering for recovery in grid graphs and planar graphs, 
and gave conditions for approximate recovery in terms of an expansion-related condition.


\paragraph{Comparison of Results.}
The two works that are most similar in the nature of guarantees are \citep{MS10} and \citep{Chenetal}. \citeauthor{MS10} designed an algorithm based on semidefinite programming (SDP relaxations with $\ell_2^2$-triangle inequality constraints) for their semi-random model on complete graphs. It finds a clustering of cost at most $1+O(n^{-1/6})$ times the cost of the optimal clustering (as long as $\varepsilon \le 1/2-O(n^{-1/3})$) and manages to approximately recover the ground truth solution (when the clusters have size at least $\sqrt{n}$). However, this algorithm only works on complete graphs and assumes unit edge costs. \citeauthor{Chenetal} studied the problem on sparser graphs from the Erd\"os--Renyi distribution, and using weaker convex relaxations gave an algorithm that recovers the ground-truth when $p \ge k^2 \log^{O(1)}n /n$. In the case of Erd\"os--Renyi graphs, our algorithms obtain similar guarantees for smaller values of $k$ (the implicit dependence on $k$ is a worse polynomial than in \citep{Chenetal}, however). The main advantage of our algorithms is that they 
work for more general graphs $G$: the first algorithm requires only that the average degree of $G$ is some poly-log of $n$, while the second algorithm requires additionally  that the graph has a mild expansion and regularity; its performance depends softly on the expansion and regularity parameters of the graph.

\subsection{Empirical Results}
This paper focuses on designing an algorithm with provable theoretical guarantees for correlation clustering in a natural semi-random model.
We have tested our algorithm to confirm that it is easily implementable and scalable.
We used the SDPNAL MATLAB library to solve the semidefinite programming (SDP) relaxation for the problem~\citep{SDPNAL}.
We implemented the recovery algorithm from Section~\ref{sec:overview} in C++, and also used a simple cleanup step that merges 
small clusters with the larger clusters based on their average inner products (this extra step can only improve our theoretical guarantees). 
We note that we could solve the SDP relaxation for instances with
 thousands of vertices since we used a very basic SDP relaxation without $\ell_2^2$-triangle inequality constraints.

We tested the algorithm on random $G(n,p)$ graphs with $4$ planted clusters of size $n/4$ each, with the error rate (the probability of flipping the label) $\varepsilon =0.2$.
We used the same values of $n$ as were used in~\citep{Chenetal}; we chose values of $p$ smaller than or close to the minimal values for which the algorithm of~\citep{Chenetal}
works (\citeauthor{Chenetal} do not report the exact values of probabilities $p$; we took approximate values from Figure~2 in their paper).
We summarize our results in Table~1. 
 \begin{table}
 \footnotesize
\begin{tabular}{|r|l|c|c|c|c|c|c|}
\hline
 & & \multicolumn{4}{|c|}{\footnotesize run number}  & &\\
\multicolumn{1}{|l|}{$n$}  & $p$ & \tiny 1 & \tiny 2 & \tiny 3 & \tiny 4 & avg. & $\%$\\
\hline
200 & $0.25$ &0 & 0 & 2 & 2 & 1 & $0.50\%$\\
\hline
400 & $0.19$ &6 & 6 & 4 & 4 & 5 &  $1.25\%$\\
\hline
1000 & $0.15$ &0 & 0 &0 &0 & 0 & $0.00\%$\\
\hline
2000 & $0.13$ &0 & 0&0&0 & 0 & $0.00\%$\\
\hline
\end{tabular}
\label{tab:exp_results}
\hspace{0.5cm}
\begin{minipage}[c]{7.7cm}
{\footnotesize  Table 1: The table summarizes results of our experiments.
The first and second columns list the values of $n$ and $p$, respectively. The next four columns list the number of misclassified
vertices in 4 runs of the program; column 7 lists the average number of misclassified vertices; column 8 shows this number as the percent of the total number of vertices.}
\end{minipage}
\end{table}
\vspace*{-1mm}
\section{Overview of the Algorithms and Structural Insights}\label{sec:overview}
\textbf{SDP relaxation.} We use a simple SDP relaxation for the problem~\citep{Swamy}. For every vertex $u$, we have a unit vector $\bar{u}$. For
two vertices $u$ and $v$, we interpret the inner product $\langle \bar u, \bar v \rangle \in [0,1]$ as the indicator of the event:
$u$ and $v$ lie in the same partition. The SDP is given below:

$$\min_{\calP}
\sum_{(u,v)\in \EE} c(u,v)(1 - \langle \bar{u}, \bar{v} \rangle) +
\sum_{(u,v)\in \EN} c(u,v)\langle \bar{u}, \bar{v}\rangle.$$
{subject to: }for all $u,v \in V$,
\vskip -2em
\begin{eqnarray*}
\langle \bar{u}, \bar{v}\rangle &\in& [0,1];\\
\|\bar u\|^2 &=& 1.
\end{eqnarray*}

The intended vector (SDP) solution has one co-ordinate for every cluster of the clustering $\calP$: the vector $\bar{u}$ for vertex $u$ has $1$ in the co-ordinate corresponding to $\calP(u)$ and $0$ otherwise. Hence this SDP is a valid relaxation.
We note that this relaxation is weaker than the SDP used in \citep{MS10} because it does not have $\ell_2^2$-triangle inequalities constraints. Hence, this semidefinite program is more scalable, and it is efficiently solvable for instances with a few thousand nodes.


\paragraph{Approximation Algorithm (PTAS).} We now describe the algorithm that gives a PTAS. Fix a parameter $\delta = o(1)\in (0,1/2)$. To simplify the notation,
denote by $f(u,v)$ (for $(u,v)\in E$) the SDP value of the edge (without cost):
\begin{equation}\label{eq:def-f}
f(u,v)=1-\uv \text{ if }(u,v) \in \EE, \text{ and } f(u,v) = \uv, \text{ otherwise.}
\end{equation}
Our PTAS is based on a surprising structural result about near-integrality of the SDP relaxation on the edges of the graph (see Theorem~\ref{thm:struct} for a formal statement).

\vspace{2mm}
\textbf{Informal Structural Theorem.}
\emph{
In any feasible SDP solution of cost at most $OPT$, the SDP value of edge $f(u,v) \ge 1-\delta $ for a $1- o_\delta(1/\log n)$ fraction of the inconsistent edges $(u,v) \in E(G)$. }
\vspace{1mm}

Hence, the structural result suggests that by removing all edges that contribute at least $(1-\delta)$ to the objective, the remaining instance has a solution of very small cost. We then run the $O(\log n)$ worst-case approximation algorithm of 
\citep{CGW} or 
\citep{DEFI} on the remaining graph to obtain a PTAS overall.

\vspace{2mm}
\textbf{Recovery.} The algorithm outlined above finds a solution of near optimal cost. Under additional assumptions, we show that we can in fact design a very simple greedy rounding scheme that can also efficiently recover the ground truth clustering approximately.

The structural theorem above shows that the SDP vectors are highly correlated for pairs of adjacent vertices.
Under the additional conditions, we show that the vectors are in fact globally clustered according to the ground truth clustering:

\paragraph{Informal Structural Theorem.}
\textit{When the semi-random instance $\Set{G=(V,E,c),\EE,\EN}$ satisfies Assumption~\ref{assumptions:recovery}, we have w.h.p. that:
for a $(1-O(\eta))$ fraction of the clusters $P_i^*$ we can choose centers $u_i\in P_i^*$ and define cores $\text{core}(P_i^*) = \Set{v\in P_i^*: \|\bar v - \bar u_i\| \leq 1/10} \subseteq P_i^*$
 (balls of radius $1/10$ around centers $\bar{u}_i$) such that
 $\core(P_i^*) \geq (1-\eta) \lvert P_i^* \rvert$ (the core of $P_i^*$ contains all but an $\eta$ fraction of vertices of $P_i^*$) and centers $u_i$
 are mutually separated by a distance  of at least $4/5$. }

The recovery algorithm is a greedy algorithm that finds heavy regions -- sets of vectors that are clumped together  -- and puts them into clusters.

\begin{tabbing}
\quad\=\quad\=\kill
 \textbf{Input:} an optimal SDP solution $\Set{\bar u}_{u\in V}$.\\[0.15em]
 \textbf{Output:} partition $P_1,\dots, P_t$ of $V$ (for some $t$).\\[0.3em]
\> $i = 1$, $\rclose = 0.1$ \\[0.15em]
\> Define an auxiliary graph $G_{aux} = (V, E_{aux})$ with $E_{aux} = \Set{(u,v): \|\bar u - \bar v\| \leq \rclose}$\\[0.15em]
\>\textbf{while} $V\setminus (P_1 \cup \dots P_{i-1}) \neq \varnothing$ \\[0.15em] 
\>\> Let $u$ be the vertex of maximum degree in $G_{aux}[V\setminus (P_1 \cup \dots P_{i-1})]$. \\[0.15em]
\>\> Let $P_i = \Set{v\notin P_1\cup \dots \cup P_{i-1}: (u,v) \in E_{aux}}$  \quad $/\mkern-6mu/$ note that $P_i$ contains $u$ \\[0.15em]
\>\>$i = i+1$ \\[0.15em]
\>\textbf{return} clusters $P_1,\dots, P_{i-1}$.
\end{tabbing}

This structural result about the global clustering and near integrality of the SDP vectors is consistent with empirical evidence. While our algorithm succeeds when the SDP is tight (as in \citep{Chenetal}), the analysis of our  algorithm also shows how to deal with nearly integral solutions, in which most inner products $\langle \bar u, \bar v\rangle$ are only close to $0$ or $1$ (but may not be tight).  We believe that many instances arising in practice have SDP solutions that are nearly integral, but not integral. Hence, we believe that in practice, our algorithm will work better than previously known algorithms.

\section{Polynomial-time Approximation Scheme}
In this section, we present the analysis of our polynomial-time approximation scheme for correlation clustering, which we presented in
Section~\ref{sec:overview}. The PTAS works in a very general Adaptive Model, which we describe first.

\subsection{Adaptive Model}\label{sec:adaptive}
We study a more general ``adaptive'' semi-random model. A semi-random instance is generated as follows. We start
with a graph $G_0(V,\varnothing)$ on $n$ vertices with no edges and a partition $\calP^*$ of $V$
into disjoint sets, which we call the planted partition. The adversary adds edges one by one. We denote
the edge chosen at step $t$ by $e_t$ and its cost $c(e_t) \in [0,1]$. After the adversary adds an edge $e_t$ to
the set of edges, the nature flips a coin and with probability $\varepsilon$ adds $e$ to the set of
random edges $E_R$. The next edge $e_{t+1}$ chosen by the adversary may depend on whether $e_t$
belongs to $E_R$ or not. The adversary stops the semi-random process at a stopping time $T$. Thus, we
obtain a graph $G^*(V,\{e_1,\dots, e_T\},c)$ and a set of random edges $E_R$. We denote the set of
all edges by $E^* = \{e_1,\dots, e_T\}$. The adversary may remove some edges belonging to $E_R$ from
the set $E^*$. Denote the set of the remaining edges by $E$. Note that
$E^*\setminus E_R\subset E\subset E^*$.

Once the graph $G(V,E)$ and the set $E_R$ are generated, we perform steps 3 and 4 from the basic semi-random model
for the graph $G(V,E)$ and random set of edges $E_R\cap E$ (as described in Section~\ref{sec:semirandom-informal}).
We obtain a semi-random instance.
This is the instance the algorithm gets. Of course, the algorithm does not get the set of random edges $E_R$.
Note that the cost of the planted solution $\calP^*$ is at most the cost of the edges $E_R\cap E$ i.e. $c\big( E_R \cap E\big)$, since all edges in $E\setminus E_R$
are consistent with $\calP^*$.

This Adaptive Model is more general than the Basic Semi-random model we introduced earlier. The basic semi-random model corresponds to the case when the whole set of edges $E^*$ is fixed in advance independent of the random choices made in $E_R$, and $E=E^*$. However, in the adaptive model the edge $e_t$ can be chosen based on which of the edges $e_1, \dots e_{t-1}$ belong to $E_R$. For instance, the adversary can choose edge $e_t$ from the portion of the graph where many of the previously chosen edges belong to $E_R$.


\subsection{Analysis of the Algorithm}
Now we analyze the algorithm presented in Section~\ref{sec:overview}.
We need to bound the number of edges removed at the first step (that is, edges $(u,v)$ with $f(u,v) > 1 -\delta$)
and the number of edges cut by the $O(\log n)$ approximation algorithm at the second step.
The SDP contribution of every edge $(u,v)$ removed at the first step is at least  $c(u,v) (1-\delta)$. Thus the cost
of edges removed at the first step is bounded by $SDP/(1-\delta)\leq (1+2\delta) OPT$. To bound
the cost of the solution produced by the approximation algorithm at the second step,
we need to bound the cost of the optimal solution for the remaining instance i.e.,
the instance with the set of edges $\{(u,v)\in E: f(u,v)\leq 1 -\delta\}$.

For any subset of edges $F \subset E$, let $c(F)$ represent the cost of the edges in $F$ i.e. $c(F)=\sum_{e \in F} c(e)$.
Denote $E^*_+$ and $E^*_{-}$: $E^*_+ = \{(u,v): \calP^*(u) = \calP^*(v)\}$ and $E^*_{-} = \{(u,v): \calP^*(u) \neq \calP^*(v)\}$.
Now define a function $f^*(u,v)$, which slightly differs from $f(u,v)$.
For all $(u,v)\in E$,
\begin{equation}
f^*(u,v) =
\begin{cases}
1 - \uv,&\text{if } \calP^*(u) = \calP^*(v);\\
\uv,&\text{if } \calP^*(u) \neq \calP^*(v).
\end{cases}\label{eq:def-f-star}
\end{equation}
Here, $\calP^*$ is the planted partition. Note that $\calP^*$ and $f^*(u,v)$ are not known to the algorithm. Observe that $f (u,v) = f^*(u,v)$ if the edge $(u,v)$ is consistent with the planted partition $\calP^*$, and  $f (u,v) = 1 - f^*(u,v)$ otherwise. Our goal is to show that the algorithm removes all but very few edges inconsistent with $\calP^*$, i.e.,
edges $(u,v)$ with  $f (u,v) = 1 - f^*(u,v)$. We prove the following theorem in Section~\ref{sec:struct}.
The proof relies on Theorem~\ref{cor:game2} presented in Section~\ref{sec:game}.

\begin{theorem}\label{thm:struct}
Let $\{G=(V,E,c)$, $(\EE, \EN)\}$ be a semi-random instance of the correlation
clustering problem. Let $E_R$ be the set of random edges, and $\calP^*$ be the planted partition.
Denote by $Q\subset E_R$ the set of random edges not consistent with $\calP^*$.
Then, for some universal constant $C$ and every $\delta, \gamma>0$, and for $\Lambda = C (1-2\varepsilon)^{-2} \gamma^{-2} \delta^{-3} n \log n $, 
$$\Pr\left[\sum_{(u,v) \in Q: f(u,v) \leq 1- \delta} c(u,v) \geq \Lambda + \frac{6\gamma}{1-2\varepsilon}c(Q)\right] = o(1).$$
where $f$ corresponds to any feasible SDP solution of cost at most $OPT$.
\end{theorem}

\begin{remark}  In the statement of Theorem~\ref{thm:struct}, $c(Q)$ is the value of the solution given by the planted solution $\calP^*$.
If $OPT=c(Q)$, then the planted solution $\calP^*$ is indeed an optimal clustering. The function $f(u,v)$ in the theorem that corresponds
to the SDP contribution of edge $(u,v)$ could come from any (not necessarily optimal) SDP solution of cost at most $OPT$. This will
be useful in Lemma~\ref{lem:lowerbound}.
\end{remark}

Let $D=O(\log n)$ be the approximation algorithms of 
\cite{CGW} or 
\cite{DEFI}. We apply Theorem~\ref{thm:struct}
with
$\gamma=\frac{\delta (1-2\varepsilon)}{6D}$.
The cost of edges in $\{(u,v) \in Q: f(u,v) \leq 1- \delta\}$ is bounded by
\begin{equation}\label{eq:boundLongQ}
\Lambda + \frac{6\gamma}{1-2\varepsilon}c(Q)\leq \Lambda + D^{-1}\delta\, c(Q),
\end{equation}
w.h.p., where $\Lambda = O((1-2\varepsilon)^{-4} \delta^{-3} n \log^3 n)$. Thus, after removing
edges with $f(u,v)\geq (1-\delta)$, the cost of the optimal solution is at most
(\ref{eq:boundLongQ}) w.h.p. The approximation algorithm finds a solution of cost
at most $D$ times (\ref{eq:boundLongQ}). Thus, the total cost of the solution returned by the algorithm
is at most 
\begin{eqnarray*}
(1+2\delta) OPT + D\times (\Lambda + D^{-1}\delta\cdot c(Q)) &=& (1+3\delta) c(Q) +  D \Lambda\\
&=&(1+3\delta)  c(Q) + O((1-2\varepsilon)^{-4} \delta^{-3} n \log^3 n).
\end{eqnarray*}

The above argument shows that the solution has small cost compared to the cost of the planted solution $\calP^*$. We can in fact use Theorem~\ref{thm:struct}
to give a true approximation i.e., compared to the cost of the optimal solution $OPT$. This follows from the following lower bound on $OPT$ in terms of $c(Q)$
for semi-random instances. 
\begin{lemma}\label{lem:lowerbound}
In the notation of Theorem~\ref{thm:struct}, with probability $1-o(1)$, 
\[ c(Q) \le (1+2\delta) OPT+O\left((1-2\varepsilon)^{-4} \delta^{-3} n \log^3 n\right).\]
\end{lemma}
\begin{proof}
Let $f_{OPT}$ correspond to the ``integral'' SDP solution corresponding to the optimal solution $OPT$. In this solution, $f_{OPT}(u,v)=1$ for positive edges $(u,v)$ which are across different clusters and negative edges $(u,v)$ which are in the same cluster. This SDP solution has cost $OPT$ and satisfies the conditions of Theorem~\ref{thm:struct}. Hence, w.h.p., $ c\left(Q \setminus (Q \cap OPT)\right) \le  \frac{\delta}{D}\cdot  c(Q) + \Lambda$. Hence,
$$c(Q)-OPT \le \frac{\delta}{D} c(Q) + \Lambda \qquad \text{and} \qquad OPT \ge (1- \frac{\delta}{D})\cdot c(Q) - \Lambda. $$
\end{proof}


We now conclude the analysis of the algorithm. 
\begin{proof}[Proof of Theorem~\ref{thm:main}]
From Theorem~\ref{thm:struct}, we get the total cost of the solution is bounded by
\begin{align*}
(1+2\delta) OPT + D\times (\Lambda + D^{-1}\delta\cdot c(Q))&=(1+2\delta) OPT + D \times \Lambda + \frac{\delta}{1- \delta/ D} \cdot (OPT+\Lambda) \\
 &\le  (1+4\delta) OPT +  2 D \Lambda \\
&=(1+4\delta) OPT + O((1-2\varepsilon)^{-4} \delta^{-3} n \log^3 n).
\end{align*}
This finishes the analysis of the algorithm.
\end{proof}
\vspace*{-0.8cm}

\subsection{Structural Theorem -- Proof of Theorem~\ref{thm:struct}}\label{sec:struct}

We now prove the Structural Theorem (Theorem~\ref{thm:struct}) assuming Theorem~\ref{cor:game2}. In order to use Theorem~\ref{cor:game2}, we need to prove that the set of all SDP solutions to our problem has a small
epsilon net. We use the following lemma from \cite{MMVfas}.

\begin{lemma}[ITCS, Lemma~2.7]\label{lem:classW}
For every graph $G=(V,E)$ on $n$ vertices ($V=\{1,\dots, n\}$) with the average degree
$\Delta = 2|E|/|V|$, real $M \geq 1$, and $\gamma \in (0,1)$, there exists a set of matrices
$\calW$ of size at most $|\calW|\leq \exp(O(\frac{n M^4\log \Delta}{2\gamma^2} + n\log n))$ such that:
for every collection of vectors $L(1),\dots, L(n)$, $R(1),\dots R(n)$ with
$\|L(u)\| = M$, $\|R(v)\| = M$ and $\langle L(u), R(v)\rangle\in [0,1]$,
there exists $W\in \calW$ satisfying for every $(u,v)\in E$:
$$ w_{uv}\leq \langle L(u), R(v)\rangle \leq w_{uv} + \gamma;$$
$$ w_{uv}\in [0,1].$$
\end{lemma}

By letting $G$ be the complete graph, $M=1$, $L(u) = R(u) = f(u)$, we get the following corollary.

\begin{corollary}\label{cor:classW}
For every $\gamma\in (0,1)$, there exists a set of matrices $\calW$ of size at most
$|\calW|\leq \exp\big(O(n \gamma^{-2}\log n)\big)$
such that:
For every collection of vectors $\{f(u)\}$, there exists $W\in \calW$ satisfying for every $(u,v)$:
$$|w_{uv} -  \langle f(u), f(v)\rangle| \leq \gamma.$$
\end{corollary}

Define $f$ and $f^*$ as in~(\ref{eq:def-f}) and (\ref{eq:def-f-star}). Recall, that the
algorithm removes all edges $(u,v)\in E$ with $f(u,v) \geq (1-\gamma)$. We show
that the number of edges inconsistent with the planted partition $\calP^*$ that
are remain in the graph after the fist step of the algorithm
is small with high probability.

\begin{proof}[Proof of Theorem~\ref{thm:struct}]
For $(u,v)\in E$, let
$$
X_{(u,v)} = \begin{cases}
1,&\text{if } (u,v) \in E_R;\\
-1,&\text{otherwise.}
\end{cases}
$$
Let $Q_+=E_R$ and $Q_{-} = E^*\setminus E_R$. Then, $Q\subset Q_+$.
Observe, that $f(u,v) = f^*(u,v)$ if $(u,v)\in E\setminus Q = Q_{-}$ and
$f(u,v) = 1 - f^*(u,v)$ if $(u,v)\in Q\subset Q_+$. The SDP value is upper bounded by the optimal value
$OPT$, which in turn is at most $c(Q)$. Write,
$$SDP = \sum_{(u,v)\in E} c(u,v) f(u,v) =  \sum_{(u,v)\in E\setminus Q} c(u,v) f^*(u,v) + \sum_{(u,v)\in Q} c(u,v) (1 - f^*(u,v)) \leq c(Q).$$
Therefore,
\begin{equation*}
\sum_{(u,v)\in E\setminus Q} c(u,v) f^*(u,v) \leq c(Q) - \sum_{(u,v)\in Q} c(u,v) (1 - f^*(u,v)) =  \sum_{(u,v)\in Q}  c(u,v) f^*(u,v).
\end{equation*}
We rewrite this expression as follows,
\begin{equation}\label{eq1}
\sum_{(u,v)\in Q \cup Q_{-}} X_{(u,v)} c(u,v) f^*(u,v)\geq 0.
\end{equation}

Suppose that
$$\sum_{(u,v) \in Q: f(u,v) \leq 1- \delta} c(u,v)  \geq \Lambda + \frac{6\gamma}{1-2\varepsilon}c(Q).$$
For $(u,v) \in Q$, $f(u,v) = 1 - f^*(u,v)$. Thus, $\{(u,v) \in Q: f(u,v) \leq 1- \delta\} =
\{(u,v) \in Q: f^*(u,v) \geq \delta\}$, and
\begin{equation}\label{eq2}
\sum_{(u,v)\in Q} c(u,v) f^*(u,v) \geq \delta \Lambda + \frac{6\delta \gamma}{1-2\varepsilon}c(Q).
\end{equation}
By Theorem~\ref{cor:game2} and Corollary~\ref{cor:classW}, the probability that inequalities~(\ref{eq1}) and (\ref{eq2})
hold is at most
$$2 \exp\big(O(n \gamma^{-2}\delta^{-2}\log n)\big)\exp\big(-\nicefrac{1}{5}(1-2\varepsilon)^2\delta\Lambda\big) = o(1),$$
for an appropriate choice of the constant $C$ in the bound on $\Lambda$.
\end{proof}

\section{Betting with Stakes Depending on the Outcome}\label{sec:game}
We first informally describe the theorem we prove in this section. Consider the following game. Assume
that we are given a set of vectors $\calW\subset [0,1]^m$. At every step $t$, the player (adversary) picks
an arbitrary not yet chosen coordinate $e_t\in\{1,\dots, m\}$, and the casino (nature) flips a coin such that
with probability $\varepsilon < 1/2$, the player wins, and with probability $(1-\varepsilon) > 1/2$, the player looses.
In the former case, we set $X_t=1$; and in the latter case we set $X_t = -1$. At some point $T \leq m$ the player
stops the game. At that point, he picks a vector $w\in \calW$ and declares that at time $t$ his stake was
$w(e_t)$ dollars. We stress that the vector $w$ may depend on the outcomes $X_t$. Then,
the player's payoff equals
$$\sum_{t=1}^T X_t w(e_t).$$
If the player could pick an arbitrary $w$ after the outcomes $X_t$ are revealed, then clearly he could
get a significant payoff by letting $w(e_t)=1$, for $X_t=1$, and $w(e_t)=0$, otherwise.
However, we assume that the set $\calW$ of possible bets is relatively small. Then, we show that with
high probability the payoff is negative unless the total amount of bets $\sum_t w(e_t)$
is very small. The precise statement of the theorem (see below) is slightly more technical.

The main idea of the proof is that for any $w\in \calW$ fixed  in advance, the player is expected to loose with high probability,
since the coin is not fair ($\varepsilon <1/2$), and thus the casino has an advantage. In fact, the probability that the player wins is exponentially small if the coordinates of $w$ are sufficiently large. Now we union bound over all $w$'s in $\calW$
and conclude that with high probability for every $w\in \calW$, the player's payoff is negative.

When we apply this theorem to a semi-random instance of Correlation Clustering (with unit costs i.e. $c(e_t)=1$), the stakes are defined by the
solution of the SDP: for an edge $e_t = (u,v)$, $w(e_t) = f^*(u,v)$. Loosely speaking, we show that since the SDP value is at
most $OPT$, the game is profitable for the adversary. This implies that
most stakes $f^*(u,v)$ are close to 0. Now, if an edge $(u,v)$ is consistent with the planted
partition $\calP^*$, then $f(u,v) = f^*(u,v)\approx 0$, and hence we do not remove
this edge. On the other hand, if the edge is not consistent with the planted partition, then
$f(u,v) = 1 - f^*(u,v)\approx 1$, hence we remove the edge.

\begin{lemma}\label{thm:game1}
Let $\calW\subset [0,1]^m$ be a set of vectors. Consider a stochastic process $(e_1,X_1,c_1),\dots, (e_T,X_T,c_T)$.
Each $e_t\in \{1,\dots, m\}\setminus \{e_1,\dots, e_{t-1}\}$, $X_t \in \{\pm1\}$, $c_t\in[0,1]$.
Let $\calF_t$ be the filtration generated by the
random variables $(e_1,X_1,c_1),\dots, (e_{t},X_{t},c_t)$, and $\calF'_t$ be the filtration generated by the
random variables $(e_1,X_1,c_1),\dots, (e_{t},X_{t},c_t)$ and $(e_{t+1},c_{t+1})$. The random variable $T\in\{1,\dots, m\}$ is a stopping time w.r.t. $\calF_t$. Each $X_t$ is a Bernoulli random variable independent of $\calF'_{t-1}$.
$$
X_t = \begin{cases}
1,&\text{with probability } \varepsilon;\\
-1,& \text{with probability } 1 - \varepsilon;
\end{cases}
$$
where $\varepsilon < 1/2$.
Then, for all $\Lambda > 3(1-2\varepsilon)^{-2}$,
\begin{multline}\label{eq:thm-game1}
\Pr\Big(\exists w \in \calW\;\text{s.t.}\; \sum_{t=1}^T X_t w(e_t) c_t+ \frac{1-2\varepsilon}{2}\sum_{t = 1}^T w(e_t) c_t\geq 0
\text { and } \sum_{t = 1}^T w(e_t) c_t \geq \Lambda \Big) \leq\\\leq 2 |\calW| e^{-\nicefrac{1}{5}(1-2\varepsilon)^2\Lambda}.
\end{multline}
\end{lemma}
\begin{proof}
To prove the desired upper bound~(\ref{eq:thm-game1}), we estimate the probability that
$\sum_{t=1}^T X_t w(e_t)c_t + \frac{1-2\varepsilon}{2}\sum_{t = 1}^T w(e_t)c_t\geq 0$
and
$\sum_{t=1}^T w(e_t)c_t\in [\Lambda',2\Lambda']$
for a fixed $w\in \calW$ and $\Lambda'\geq \Lambda$. Then we apply the union bound
for all $w\in \calW$, and $\Lambda'$ of the form $2^i\Lambda$.

Fix a $w\in \calW$ and $\Lambda'=2^i$. Each $X_{t+1}$ is independent of $\calF'_t$, hence
$\Exp[X_{t+1} w(e_{t+1}) c_{t+1}\given \calF'_t] = \Exp[X_{t+1}] w(e_{t+1})c_{t+1} = (2\varepsilon - 1)w(e_{t+1})c_{t+1}$.
Thus,
$$S_{\tau} \equiv \sum_{t=1}^{\tau} (X_{t} + 1 -2 \varepsilon)  w(e_{t})c_{t}$$
is a martingale.
Note that $|S_{t+1} - S_t|\leq w(e_{t+1})c_{t+1}\leq c_{t+1}$ and
$$\Var[X_{t+1} w(e_{t+1}c_{t+1})\given \calF'_t] = 4\varepsilon(1 -\varepsilon) w(e_{t+1})^2 c_{t+1}^2\leq 4\varepsilon(1 -\varepsilon) w(e_{t+1})c_{t+1}.$$

\medskip

If $\sum_{t=1}^T X_t w(e_t)c_{t} + \frac{1-2\varepsilon}{2}\sum_{t = 1}^T w(e_t)c_{t}\geq 0$ and $\sum_{t=1}^T w(e_t)c_{t}\in [\Lambda',2\Lambda']$,
then
$$S_{T} = \left[\sum_{t=1}^T X_t w(e_t)c_{t} + \frac{(1-2\varepsilon)}{2} \sum_{t=1}^T w(e_t)c_{t}\right] + \frac{(1-2\varepsilon)}{2} \sum_{t=1}^T w(e_t)c_{t}\geq \frac{(1-2\varepsilon)}{2} \Lambda',$$
and
$$\sum_{t=1}^T \Var[X_{t} w(e_{t})c_{t}\given \calF'_{t-1}] = 4(\varepsilon -\varepsilon^2) \sum_{t=1}^T  w(e_{t})c_{t}
\leq 8\varepsilon (1-\varepsilon)\Lambda'.$$
Now, by Freedman's inequality (see~\cite{freedman}),
\begin{align*}
\Pr\Big(S_T \geq (1-2\varepsilon)\Lambda' \;\text{and}\;
\sum_{t=1}^T \Var[X_{t} w(e_{t})c_{t}\given \calF_{t-1}]\leq  8\varepsilon (1-\varepsilon)\Lambda' \Big)&\leq e^{-\frac{(1-2\varepsilon)^2\Lambda'^2}{2((1-2\varepsilon)\Lambda' + 8\varepsilon (1-\varepsilon)\Lambda')}}\\&=
e^{-\frac{(1-2\varepsilon)^2\Lambda'}{5}},
\end{align*}
and
\begin{eqnarray*}
\Pr\Big(\sum_{t=1}^T X_t w(e_t)c_{t} \geq 0\;\text{and}\;
\sum_{t=1}^T w(e_t)c_{t}\in [\Lambda',2\Lambda']\Big) &\leq&
\Pr\Big(S_T \geq (1-2\varepsilon)\Lambda' \;\text{and}\;
\sum_{t=1}^T w(e_t)^2c_{t}^2\leq 2\Lambda'\Big)\\
&\leq& e^{-\nicefrac{1}{5}\,(1-2\varepsilon)^2\Lambda'} =
\big(e^{-\nicefrac{1}{5}\,(1-2\varepsilon)^2\Lambda})^{2^i}.
\end{eqnarray*}
Summing up this upper bound over all $w\in \calW$ and $\Lambda' = 2^i\Lambda$, we get~(\ref{eq:thm-game1}).
\end{proof}

We now slightly generalize this theorem. In our application, the set of all possible stakes can be infinite, however,
we know that there is a relatively small epsilon net for it.

\begin{definition}
We say that a set $\calW\subset\bbR^m$ is a $\gamma$--net for a set $\calZ\subset \bbR^m$ in the $\ell_{\infty}$ norm, if for every
$z\in \calZ$, there exists $w\in \calW$ such that
$\|z-w\|_{\infty}\equiv \max_i \{|z(i) - w(i)|\}\leq \gamma$.
\end{definition}

\begin{remark}\label{rem:truncateNet}
If $\calW$ is a $\gamma$--net for $\calZ\subset [0,1]^m$, then there exists $\calW'\subset [0,1]^m$ of the same size as $\calW$
($|\calW'|=|\calW|$), such that for every $z\in \calZ$, there exists $w'\in \calW'$ satisfying $w'(i) \leq z(i)\leq w'(i) + 2\gamma$
for all $i$.
To obtain $\calW'$ we simply subtract $\min(\gamma, w(i))$ from each coordinate of $w$ and then truncate each $w'(i)$ at the threshold of 1.
\end{remark}

\begin{theorem}\label{cor:game2}
Consider a stochastic process $(e_1,X_1, c_1),\dots, (e_T,X_T,c_T)$ such that each $e_t\in \{1,\dots, m\}\setminus \{e_1,\dots, e_{t-1}\}$, $X_t\in\{\pm 1\}$ and $c_t\in[0,1]$.
Let $\calF_t$ be the filtration generated by the
random variables $(e_1,X_1,c_1),\dots, (e_{t},X_{t},c_t)$, and $\calF'_t$ be the filtration generated by the
random variables $(e_1,X_1, c_1),\dots, (e_{t},X_{t}, c_t)$ and $(e_{t+1},c_{t+1})$.  The random variable $T\in\{1,\dots, m\}$ is a stopping time
w.r.t. $\calF_t$. Each $X_t$ is a Bernoulli random variable independent of $\calF'_{t-1}$.
$$
X_t = \begin{cases}
1,&\text{with probability } \varepsilon;\\
-1,& \text{with probability } 1 - \varepsilon;
\end{cases}
$$
where $\varepsilon < 1/2$. Let $\calZ\subset [0,1]^m$
be a set of vectors having a $\gamma$--net in the $L_{\infty}$ norm of size $N$. Define two random sets depending on $\{X_t\}$:
$$Q_+ = \{t: X_t = 1\}\;\;\text{and}\;\;Q_- = \{t: X_t = -1\}.$$
Then, for all $\Lambda> 3(1-2\varepsilon)^2$, we have
\begin{multline}\label{eq:thm-game2}
\Pr\Big(\exists z \in \calZ,\;Q_{\oplus}\subset Q_+\;\text{s.t.}\; \sum_{t\in Q_{\oplus} \cup Q_-} X_t z(e_t) c_t \geq 0\\
\text { and } \sum_{t\in Q_{\oplus}} z(e_t)c_t \geq \Lambda + \frac{6 \gamma}{1-2\varepsilon}\; \sum_{t \in Q_{\oplus}} c_t \Big)
\leq 2 N e^{-\nicefrac{1}{5}(1-2\varepsilon)^2\Lambda}.
\end{multline}
\end{theorem}
\begin{proof}
Let $\calW$ be a $\gamma$--net for $\calZ$. For simplicity of exposition we subtract $\min(\gamma, w(i))$ from all coordinates
of vectors $w\in \calW$. Thus, we assume that for all $z\in \calZ$, there exists $w\in \calW$ such that $w(i) \leq z(i)\leq w(i) + 2\gamma$
and $w(i)\geq 0$ for all $i$ (see Remark~\ref{rem:truncateNet}).

Suppose that for some $z\in \calZ$ and $Q_{\oplus}\subset Q_+$, the inequalities
\begin{equation}\label{eq:sumXZ}
\sum_{t\in Q_{\oplus} \cup Q_{-}} X_t z(e_t) c_t \geq 0
\end{equation}
and
\begin{equation}\label{eq:sumZ}
\sum_{t\in Q_{\oplus}} z(e_t) c_t \geq \Lambda + \frac{6 \gamma}{1-2\varepsilon} \sum_{t \in Q_{\oplus}} c_t
\end{equation}
hold. Pick a $w\in \calW$, such that $w(i)\leq z(i)\leq w(i)+2\gamma$ for all $i$. We replace $z(e_t)$ with $w(e_t)$
in (\ref{eq:sumZ}):
\begin{equation}\label{eq:sumW}
\sum_{t\in Q_{\oplus}} w(e_t) c_t \geq \sum_{t\in Q_{\oplus}} (z(e_t) - 2\gamma) c_t \geq \Lambda +
\frac{4\gamma}{(1-2\varepsilon)} \cdot \sum_{t \in Q_{\oplus}} c_t.
\end{equation}
Then,
\begin{eqnarray}\label{eq:sumXW}
\sum_{t=1}^T X_t w(e_t) c_t +
\frac{1-2\varepsilon}{2}\; \sum_{t=1}^T w(e_t) c_t &\geq&
\sum_{t\in Q_{\oplus} \cup Q_{-}} X_t w(e_t) c_t +
\frac{1-2\varepsilon}{2}\; \sum_{t\in Q_{\oplus}} w(e_t) c_t \\
&\geq& \left[\sum_{t\in Q_{\oplus}} (z(e_t) - 2\gamma) c_t -
\sum_{Q_{-}} z(e_t) c_t \right] + 2\gamma \sum_{t \in Q_{\oplus}} c_t \notag \\
&=&\sum_{t\in Q_{\oplus} \cup Q_{-}} X_t z(e_t) c_t \geq 0.\notag
\end{eqnarray}

By Lemma~\ref{thm:game1}, there exists a $w\in \calW$ satisfying (\ref{eq:sumW}) and (\ref{eq:sumXW}) with probability
at most $2Ne^{-\nicefrac{1}{5}(1-2\varepsilon)^2\Lambda}$. This concludes the proof.
\end{proof}

\section{Recovery Algorithm}
In this section, we prove Theorem \ref{thm:recovery_algorithm} that shows that under some additional assumptions on the graph $G$ and partition $\calP^*$, we can recover the planted partition $\calP^*$ with an arbitrarily small classification error $\eta$. The recovery algorithm is a very fast and very simple greedy algorithm (presented in Section~\ref{sec:overview}).
\vspace{-2mm}
\begin{assumptions}\label{assumptions:recovery} 
Consider a semi-random instance $\calI = \Set{G = (V, E, c),(\EE, \EN)}$. Let $\calP^*$ be the planted partition. Denote the clusters of $G$ w.r.t clustering $\calP^*$ by $P_1^*, \dots, P_k^*$.
Let $\beta = c(\EE^*) / c(E)$ (note that $\EE^*$ is the set of edges that lie within clusters) and $\beta_{ij} = c(\{(u,v): u \in P_i^*, v\in P_j^*\}) / c(E)$ (here, $\{(u,v): u \in P_i^*, v\in P_j^*\}$ is the set of edges between clusters $P_i^*$ and $P_j^*$).
Assume that the instance $\calI$ satisfies the following conditions:
\begin{itemize}
\vspace{-2mm}
\item\textbf{Cluster Expansion.} All induced graphs $G[P_i^*]$ are spectral expanders with spectral expansion at least $\lgap$; that is, the second smallest eigenvalue of the normalized Laplacian of $G[P_i^*]$
is at least $\lgap$.
\vspace{-2mm}
\item\textbf{Intercluster Density.} For some sufficiently large constant $C_1$, and every two clusters $P_i^*$ and $P_j^*$,
$\beta_{ij}  >   \frac{C_1}{(1-2\varepsilon)^2} \left(\frac{n\log n}{c(E)}\right)^{1/6}$.
\vspace{-2mm}
\item\textbf{Intercluster Regularity.} The set of edges between every two clusters $P_i^*$ and $P_j^*$ forms a regular graph with respect to the cost function $c$: for every $u', u'' \in P_i^*$ we have
$c(\{(u',v): v\in P_j^*\}) = c(\{(u'',v): v\in P_j^*\})$.
\vspace{-2mm}
\item \textbf{Cluster Regularity.} All induced graphs $G[P_i^*]$ are regular graphs with the same degree w.r.t to the cost function $c$. That is, for some number $c_0$, every cluster $P_i^*$,  and every vertex $u\in P_i^*$,
$c_0 = c(\Set{(u,v) \in E: v\in P_i^*})$.
\end{itemize}
\end{assumptions}
\noindent \textbf{Remark} The Intercluster and Cluster Regularity assumptions can be significantly relaxed; in fact, we only need that degrees are equal up to some multiplicative factor (say, a poly-log factor).
We include the regularity assumptions to simplify the exposition.
\begin{definition}
Let $\calI =  \Set{G = (V, E, c),(\EE, \EN)}$ be a semi-random instance of correlation clustering, $\calP^*$ be the planted partition, and $P_1^*, \dots, P_k^*$ be the planted clusters.
We say that a partition $\calP$ of $V$ into clusters $P_1,\dots, P_t$ has an $\eta$ classification error if there is a partial matching between
clusters $P_1^*, \dots, P_k^*$ and clusters $P_1,\dots, P_t$ such that
$$\sum_{P_i^* \text{ is matched with } P_j} |P_i^* \cap P_j| \geq (1 - \eta) |V|.$$
\end{definition}

Theorem~\ref{thm:recovery_algorithm} relies on the following theorem that describes the structure of optimal SDP solutions to semi-random instances of correlation clustering
that satisfy conditions in Assumption~\ref{assumptions:recovery}.

\begin{theorem}\label{thm:recovery}
Assume that a semi-random instance $\calI = \Set{G = (V, E, c),(\EE, \EN)}$ satisfies Assumptions~\ref{assumptions:recovery}.
Let $\Set{\bar u}$ be the optimal SDP solution to $\calI$.
With probability $1- o(1)$, there exist a subset of clusters $\calC\subset\Set{P_1^*,\dots, P_k^*}$ and a vertex $u_i$ in each cluster $P_i^*$ satisfying the following properties.
Let $\rclose = 1/10$ and $\rfar = 4/5$.
Let $\core(P_i^*) = \Set{v\in P_i^*: \|\bar v - \bar u_i\| \leq \rclose}$ for $P_i^* \in \calC$, then
\begin{enumerate}
\item $|\cup_{P_i^* \in \calC} P_i^*| \geq (1-\eta) |V|$.
\item $|\core(P_i^*)| \geq (1 - \eta) |P_i^*|$.
\item In particular, $\sum_{P_i^* \in \calC} |P_i| \geq (1 - \eta)^2 |V|$.
\item $\|\bar u_i - \bar u_j\| \geq \rfar$ for every two distinct clusters $P_i^*, P_j^* \in \calC$.
\end{enumerate}
\end{theorem}


\medskip

We now use Theorem~\ref{thm:recovery} to prove the recovery guarantees of our algorithm. 
\begin{proof}[Proof of Theorem~\ref{thm:recovery_algorithm}]
Consider a cluster $P_i$. Let $u$ be the vertex we choose at iteration $i$ of the while--loop.
If $P_i$ intersects a core $\core(P_{j}^*)$ of a cluster $P_j^*$ then $\|u - u_j\| \leq 2 \rclose$. Note that
$P_i$ cannot intersect cores $\core(P_{j'}^*)$ and $\core(P_{j''}^*)$ of
two distinct clusters $P_{j'}^*$ and $P_{j''}^*$, since
$\|u - u_{j'}\| +\|u - u_{j''}\|  \geq \|u_{j'} - u_{j'}\| \geq \rfar > 4\rclose$. Thus each cluster $P_i$ intersects at most the core of one cluster $P_j^*$.

We match every cluster $P_j^* \in \calC$ to the first cluster $P_i$ that intersects $\core(P_j^*)$.
Consider a cluster $P_j^* \in \calC$ and the matching cluster $P_i$. Since
$\core(P_j^*) \cap (P_1 \cup \dots \cup P_{i-1}) = \varnothing$, we have, in particular, that $u_j \notin P_1 \cup \dots \cup P_{i-1}$ and
$u_j$ has degree at least $|\core(P_j^*)|$ in $G_{aux}[V\setminus (P_1 \cup \dots P_{i-1})]$.
Thus the vertex $u$ that we choose at iteration $i$ has degree at least $|\core(P_j^*)|$ and  $|P_i| \geq |\core(P_j^*)|$; in particular,
$|P_i\setminus \core(P_j^*)| \geq |\core(P_j^*)\setminus P_i|$. We have,
$$|P_j^*  \cap P_i| \geq |\core(P_j^*) \cap P_i| = |\core(P_j^*)| - |\core(P_j^*) \setminus P_i| \geq |\core(P_j^*)| - |P_i\setminus \core(P_j^*)|.$$
Note that by Theorem~\ref{thm:recovery} (item 3)
$$\sum_{P_i \text{ is matched with } P_j^*} |P_i\setminus \core(P_j^*)| \leq |V| - |\bigcup_{P_j^* \in \calC} \core(P_j^*)|  \leq n - (1 -\eta)^2 n \leq 2\eta n.$$
Therefore,
$$\sum_{P_i \text{ is matched with } P_j^*} |P_i \cap P_j^*| \geq \Bigl(\sum_{P_j^*\in \cal C} |\core(P_j^*)|\Bigr) - 2\eta n \geq (1-\eta)^2 n - 2\eta \geq (1- 4\eta) n.$$
We proved that the algorithm finds a clustering with classification error at most $4\eta$.
\end{proof}

\subsection{Structure of the optimal SDP solution: Proof of Theorem~\ref{thm:recovery}} \label{sec:recovery}
%
%
We now prove Theorem~\ref{thm:recovery} which gives the structure of optimal SDP solutions to semi-random instances of correlation clustering that satisfy conditions in Assumption~\ref{assumptions:recovery}. As seen earlier, completing this proof concludes the proof of Theorem~\ref{thm:recovery_algorithm}.

Let  $\delta = \gamma =  (n \log n / c(E))^{1/6}$.
Let $\Lambda$ and $Q$ be as in Theorem~\ref{thm:struct}.
Let $\sigma = 6\delta/ (1-2\varepsilon)$. Note that $\Lambda = O(\delta/ (1 - 2\varepsilon)^2)$.

Define $f$ as in (\ref{eq:def-f}):
\begin{equation}
f(u,v) =
\begin{cases}
1 - \uv,&\text{if } (u,v)\in \EE;\\
\uv,&\text{if } (u,v)\in \EN.
\end{cases}
\end{equation}
Consider the set of edges $\Efix = \Set{(u,v)\in E: f(u,v) > 1 - \delta}$.
Change the sign of each edge in $\Efix$ and obtain a new partitioning of $E$  into positive and negative edges, $\hEE$ and $\hEN$:
\begin{align*}
\hEE &= \EE \symdiff \Efix = \Set{(u,v)\in \EE: f(u,v) \leq 1 - \delta} \cup \Set{(u,v)\in \EN: f(u,v) > 1 - \delta},
\\
\hEN &= \EN \symdiff \Efix = \Set{(u,v)\in \EN: f(u,v) \leq 1 - \delta} \cup \Set{(u,v)\in \EE: f(u,v) > 1 - \delta}.
\end{align*}
Let us now consider the corresponding instance
$\hcalI = \Set{G = (V, E, c),(\hEE, \hEN)}$.
Let $\hf$ be the analog of function $f$ for $\hcalI$:
\begin{equation}\label{eq:def-hf}
\hf(u,v) =
\begin{cases}
1 - \uv,&\text{if } (u,v)\in \hEE\\
\uv,&\text{if } (u,v)\in \hEN
\end{cases}
\quad = \quad
\begin{cases}
f(u,v),&\text{if } (u,v) \notin \Efix;\\
1 - f(u,v),&\text{if } (u,v)\in\Efix.
\end{cases}
\end{equation}
Similarly,  let $\hSDP = \sum_{(u,v)\in E} c(u,v) \hf(u, v)$ be the cost of the SDP solution $\Set{\bar u}$ for $\hcalI$.
\begin{lemma}\label{lem:basic-recovery}
With probability $1 - o(1)$, the following properties hold.
\begin{enumerate}
\item $c(Q \setminus \Efix) \leq \sigma c(Q) +\Lambda$.
\item $c(\Efix \setminus Q) \leq (2\delta + \sigma) c(Q) + \Lambda$.
\item Then $\hSDP \leq (2\delta + \sigma) c(Q) + \Lambda$.
\end{enumerate}
\end{lemma}
\begin{proof}
\noindent 1. From  Theorem~\ref{thm:struct}, we get that $c(Q \setminus \Efix) \leq \sigma c(Q) + \Lambda$ with probability $1 - o(1)$.

\noindent 2. Write $c(\Efix \setminus Q) = c(\Efix) - c(Q \cap \Efix)$. Now we bound $c(\Efix)$ and $c(Q \cap \Efix)$.
Note that
$$SDP = \sum_{(u,v) \in E} c(u,v) f(u,v) \geq \sum_{(u,v) \in \Efix} c(u,v) (1 - \delta) = (1 - \delta) c(\Efix).$$
Hence,
$$
c(\Efix) \leq SDP / (1 - \delta) \leq  c(Q)/ (1 -\delta) \leq (1 + 2\delta) c(Q),
$$
here, we used that $\Set{\bar u}$ is an optimal SDP solution and therefore $SDP \leq c(Q)$.

By item 1, $c(Q \cap \Efix)  = c(Q) - c(Q \setminus \Efix) \geq (1 - \sigma) c(Q) - \Lambda$.
We get that $$c(\Efix \setminus Q) \leq   (1 + 2 \delta) c(Q) - (1 - \sigma) c(Q) - \Lambda = (2\delta + \sigma) c(Q) + \Lambda.$$

\noindent 3. From the second formula for $\hf(u,v)$ in (\ref{eq:def-hf}), we get that $f(u,v) - \hf(u,v) =  2f(u,v) - 1 \geq 1 - 2\delta$ for $(u,v)\in \Efix$, and $f(u,v) - \hf(u,v) = 0$
for $(u,v) \notin \Efix$. Therefore,
\begin{align*}
c(Q) - \hSDP &\geq SDP - \hSDP = \sum_{(u,v) \in E} c(u,v) (f(u,v) - \hf(u,v)) \\
&= \sum_{(u,v) \in \Efix}  c(u,v) (f(u,v) - \hf(u,v)) \geq (1-2\delta) c(\Efix) \geq (1-2\delta) c(Q \cap \Efix) \\
&\geq (1 - 2\delta) ((1 - \sigma) c(Q) - \Lambda) \geq (1 - 2\delta - \sigma) c(Q) -  \Lambda.
\end{align*}
Therefore, $\hSDP \leq (2\delta + \sigma) c(Q) + \Lambda$.
\end{proof}

We now bound the total squared Euclidean length of all edges in $\EE^*$.
\begin{lemma} With probability $1 - o(1)$, we have
$$\frac{1}{2} \sum_{(u,v) \in \EE^*} c(u,v) \|\bar u - \bar v\|^2 \leq (4\delta + 3\sigma) c(Q) + 3\Lambda
$$
\end{lemma}
\begin{proof}
Note that for $(u,v) \in \hEE$, $\frac{1}{2}\|\bar u - \bar v\|^2 = \hf(u,v)$ and thus
$$\frac{1}{2} \sum_{(u,v) \in \hEE}  c(u,v) \|\bar u - \bar v\|^2 \leq \hSDP.$$
Also, $\EE^* \cap \hEN \subset (Q \setminus \Efix) \cup  (\Efix \setminus Q)$. Thus, by Lemma~\ref{lem:basic-recovery},
$c(\EE^* \cap \hEN) \leq c(Q \setminus \Efix) + c(\Efix \setminus Q) \leq 2(\delta + \sigma) c(Q) + 2\Lambda$.
We have,
\begin{align*}
\frac{1}{2} \sum_{(u,v) \in \EE^*} c(u,v) \|\bar u - \bar v\|^2 &\leq \frac{1}{2} \sum_{(u,v) \in \EE^* \cap \hEE} c(u,v) \|u - v\|^2
+ \frac{1}{2} \sum_{(u,v) \in \EE^* \cap \hEN} c(u,v) \|u - v\|^2 \\
& \leq \hSDP + c(\EE^* \cap \hEN) = (4\delta + 3\sigma) c(Q) + 3\Lambda.
\end{align*}
\end{proof}
We are ready to prove Theorem~\ref{thm:recovery}. 
\begin{proof}[Proof of Theorem~\ref{thm:recovery}]

We assume that $\eta < 1/4$ as otherwise the statement of theorem is trivial.
Let
$$\rsavg =  \frac{1}{c(\EE^*)}\sum_{(u,v) \in \EE^*} c(u,v) \|\bar u - \bar v\|^2 \leq O(\sigma c(Q) + \Lambda)/c(\EE^*) \leq O(\sigma + \Lambda/c(E))/\beta .$$
Let $\EE^*(i) = \Set{(u,v) \in \EE^*: u,v \in P_i^*}$ be the set of edges within cluster $P_i^*$.
Write
$$\sum_{i=1}^k \sum_{(u,v) \in \EE^*(i)} c(u,v) \|\bar u - \bar v\|^2 = c(\EE^*) \rsavg.$$
Let $\calC$ be the set of clusters $P_i^*$ such that $\sum_{(u,v) \in \EE^*(i)} c(u,v) \|\bar u - \bar v\|^2 \leq c(\EE^*(i)) \rsavg/\eta$.
By Markov's inequality, $\sum_{P_i^* \in \calC} c(\EE^*(i)) \geq (1 -\eta) c(\EE^*)$.
By the Cluster Regularity condition in Assumptions~\ref{assumptions:recovery},
 $c(\EE^*(i)) =  (|P_i^*| /n) c(\EE^*)$. We get that
  $\sum_{P_i^* \in \calC} |P_i| \geq (1 - \eta) n$ and item 1 in the statement of the theorem holds.

By the Poincar\'e inequality\footnote{Recall that the Poincar\'e inequality states that for every every expander graph $H = (V_H, E_H, c_H)$ with spectral expansion $\lambda$ and every set of vectors $\{\bar u \}_{u\in V_H}$, we have $\frac{1}{|V_H|^2}\sum_{u,v\in V} \|\bar u - \bar v\|^2 \leq \frac{1}{\lambda\cdot c_H(E_H)} \sum_{(u,v) \in E_H} c_H(u,v)\|\bar u - \bar v\|^2$. Here, we apply the Poincar\'e inequality to the induced graph $G[P_i^*]$}
we have for each cluster $P_i^*\in \calC$,
$$\frac{1}{|P_i^*|^2} \sum_{u,v\in P_i^*} \|\bar u - \bar v\|^2 \leq \frac{1}{\lgap}
\frac{1}{c(\EE^*(i))}\sum_{(u,v) \in \EE^*(i)} c(u,v) \|\bar u - \bar v\|^2 \leq \frac{\rsavg}{\lgap\eta}.$$
Therefore,
$$ \min_{u\in P_i^*} \left(\frac{1}{|P_i^*|} \sum_{v\in P_i^*} \|\bar u - \bar v\|^2 \right)\leq
\frac{1}{|P_i^*|} \sum_{u\in P_i^*} \left(\frac{1}{|P_i^*|}\sum_{v\in P_i^*} \|\bar u - \bar v\|^2  \right)\leq \frac{\rsavg}{\lgap\eta}.$$
Thus we can choose $u_i$ in each $P_i^*\in \calC$ such that $\frac{1}{|P_i^*|} \sum_{v\in P_i^*} \|\bar u_i - \bar v\|^2 \leq \frac{\rsavg}{\lgap\eta}$.
This choice of vertices $u_i$ defines sets $\core(P_i^*)$, as in the statement of the theorem.
Using again Markov's inequality, we get that for at least a $1-\eta$ fraction of vertices $v$ in $P_i^*$, $\|\bar u_i - \bar v\|^2 \leq \rsavg/(\lgap\eta^2)$.
From the bound $\rsavg = O(\sigma + \Lambda/c(E))/\beta$ and formula~\ref{eq:eta-def}, we get $\rsclose \geq \rsavg/(\lgap\eta^2)$ and
$$|\core(P_i^*)| \geq |\Set{v\in P_i^*: \|\bar u_i - \bar v\|^2 \leq \rsavg/(\lgap\eta^2)}| \geq (1 - \eta) |P_i^*|.$$
We showed that item 2 in the statement of the theorem holds. We get item 3 from items 1 and 2.

Finally, we show that $\|\bar u_i -  \bar u_j\| \geq \rfar$ for every two distinct clusters $P_i^*, P_j^* \in \calC$.
To this end, we show that there are vertices $v' \in \core(P_i^*)$ and $v'' \in \core(P_j^*)$ such that $\|\bar v' - \bar v''\| \geq \rfar + 2\rclose$, and thus
$\|\bar u_i -  \bar u_j\| \geq (\rfar + 2\rclose) - \|u_i - v'\| - \|u_j - v''\| \geq \rfar$.
Assume to the contrary that $\|\bar v' - \bar v''\| < \rfar + 2\rclose$ for every $v' \in \core(P_i^*)$ and $v'' \in \core(P_j^*)$. Let
$E_{ij} = \Set{(v',v'')\in E:v'\in \core(P_i^*), v''\in \core(P_j^*)}$.

Since $E_{ij} \subset \EN^*$, we have for every $(v',v'') \in E_{ij} \setminus (Q \symdiff \Efix)$,
$$\hf(v',v'') = \langle \bar v', \bar v''\rangle = 1 - \|\bar v' - \bar v''\|^2/2 \geq 1 - (\rfar + 2\rclose)^2/2 = 1/2.$$
Therefore,
$$\hSDP \geq \sum_{(v',v'') \in E_{ij} \setminus (Q \symdiff \Efix)} c(v',v'') f(v',v'') \geq c(E_{ij} \setminus (Q \symdiff \Efix)) /2.$$
From the Intercluster Regularity condition and bounds $|\core(P_i^*)| \geq (1-\eta) |P_i^*|$ and $|\core(P_j^*)| \geq (1-\eta) |P_j^*|$, we get
 $c(E_{ij}) \geq (1-2\eta) \beta_{ij} c(E) $ .
By Lemma~\ref{lem:basic-recovery},
$$c(Q \symdiff \Efix) \leq 2(\delta + \sigma) c(Q) + 2\Lambda \leq 2(\delta + \sigma) c(E) + 2\Lambda.$$
By the Intercluster Density condition in Assumptions~\ref{assumptions:recovery} and our choice of $\delta$, we have
$$c(E_{ij} \setminus (Q \symdiff \Efix)) \geq  ((1-2\eta) \beta_{ij}  - 2\delta - 2\sigma) c(E)- 2\Lambda \geq \beta_{ij} c(E) /3.$$
We get that
$$(2\delta + \sigma) c(Q) + \Lambda \geq \hSDP \geq \beta_{ij} c(E) /6,$$
which contradicts to the Intercluster Density condition and our choice of $\delta$.
\end{proof}

\appendix
\bibliographystyle{plainnat}
\bibliography{semirandom}
\end{document}